\documentclass[preprint,aps,nofootinbib]{revtex4}
\usepackage{amsfonts,amsmath,amssymb,amsthm}
\usepackage{latexsym}
\usepackage{bbm,bm}
\usepackage{graphicx}
\newcommand{\ket}[1]{\lvert #1 \rangle}
\newcommand{\bra}[1]{\langle #1 \lvert}
\newcommand{\beq}{\begin{equation}}
\newcommand{\eeq}{\end{equation}}
\newcommand{\beqs}{\begin{eqnarray}}
\newcommand{\eeqs}{\end{eqnarray}}
\begin{document}

\title{Mixedness and Entanglement in the presence of Localized Closed Timelike Curves}

\author{Eylee Jung$^1$ and DaeKil Park$^{1,2}$}

\affiliation{$^1$Department of Electronic Engineering, Kyungnam University, Changwon,
631-701, Korea \\
$^2$ Department of Physics, Kyungnam University, Changwon,
631-701, Korea}

\begin{abstract}
We examine mixedness and entanglement of the chronology-respecting (CR) system assuming that quantum mechanical closed timelike
curves (CTCs) exist in nature. In order discuss these two issues analytically, we introduce the qubit system and 
and apply the general controlled operations between CR and CTC systems. We use the 
magnitude of Bloch vector as a measure of mixedness. While Deutschian-CTC (D-CTC) either preserves or decreases the magnitude, 
postselected-CTC (P-CTC) can increases it. Nonintuitively, even the completely mixed CR-qubit can be converted into a pure state 
after CTC-qubit travels around the P-CTC.
It is also shown that while D-CTC cannot increase the entanglement of CR system, P-CTC can increase it. This means that any partially entangled 
state can be maximally entangled pure state if P-CTC exists. Thus, distillation of P-CTC-assisted entanglement can be easily achieved 
without preparing the multiple copies
of the partially entangled state.
\end{abstract}

\maketitle

\section{Introduction.}
It is well-known that the theory of general relativity allows the possible existence of closed timelike 
curves (CTCs)\cite{godel49,morris88,gott91}. 
However, allowance of 
time travel generates logical paradoxes such as the {\it grandfather paradox}, in which the time traveller performs an action that causes 
her future self not to exist. Furthermore, the existence of CTCs cannot be incompatible with a standard quantum mechanics, because 
quantum mechanics allows only unitary evolution. In order to solve these difficulties Deutsch\cite{deutsch91} modifies the standard 
quantum mechanics, which allows the non-unitary evolution in the presence of CTCs. To escape the {\it grandfather paradox}, in addition, 
he imposes the self-consistent constraint of CTC interaction (see Eq. (\ref{self}). Thus, it makes it possible to explore the properties of CTCs without relying on the exotic spacetime geometries.

Then, it is natural to ask how quantum mechanics is modified if Deutsch's CTCs (D-CTCs) exist. For last few years this question was explored 
in the various contexts\cite{brun03,bacon04,aaronson09,brun09,pati2011,ahn13,brun13}. Among them most striking result is that any non-orthogonal states 
can be perfectly distinguished if one can access to D-CTCs\cite{brun09}. This fact implies that Security of usual quantum cryptography scheme such as 
BB84 protocol\cite{bb84} is not guaranteed. Subsequently, the authors of Ref.\cite{bennett09} raised a question on the perfect discrimination and
computational power in the presence of D-CTCs. They argued that when the input state is a labeled mixture, the assistance of CTCs in distinguishability and
computational power is of no use. However, their argument was also criticized in Ref.\cite{ralph2010}. The authors of  Ref.\cite{ralph2010} claimed
by constructing the equivalent circuit that the CTCs would be a true powerful resource for quantum information processing.
Another nonintuitive result arising due to existence of D-CTCs is that any arbitrary dimensional quantum states 
can be perfectly cloned if the dimension of the CTC system is infinite\cite{ahn13,brun13}. Thus, the well-known no-cloning theorem\cite{noclonning} can
be broken in the presence of D-CTCs. In our opinion, however, still it seems to be open problem to determine whether or not the assistance of CTCs
allows such nonintuitive results, because the debate between Ref.\cite{brun09} and Ref.\cite{bennett09} is not concluded yet.

Mathematically, the Deutsch's self-consistency condition is expressed as 
\begin{equation}
\label{self}
\rho_{out}^{(CTC)} \equiv \mbox{tr}_{CR} \left[ U \left(\rho_{in}^{(CR)} \otimes \rho_{in}^{(CTC)} \right) U^{\dagger} \right] = \rho_{in}^{(CTC)}
\end{equation}
where $\rho_{in}^{(CR)}$ and $\rho_{in}^{(CTC)}$ are input states of the chronology-respecting (CR) and  chronology-violating systems, respectively.
Here, $\rho^{(CTC)}$ is a quantum state of system traversing the CTC and $\rho^{(CR)}$ is a quantum state of system, 
which only interacts with  $\rho^{(CTC)}$, but not  traversing the CTC. 
The operator $U$ represents the unitary interaction between CR and CTC systems. Since the self-consistency condition imposes the 
equality of input and output CTC states, it naturally solves the {\it grandfather paradox}.
Deutsch showed\cite{deutsch91} that the fixed-point solution of Eq. (\ref{self}) always exists, but it does not necessarily have to be unique. If there are many solutions, Deutsch suggested the {\it maximum entropy rule}.
If $\rho^{(CTC)}$ is fixed, the CR system is evolved as 
\begin{equation}
\label{CRevolution}
\rho_{in}^{(CR)} \rightarrow \rho_{out}^{(CR)} \equiv \mbox{tr}_{CTC} \left[ U \left(\rho_{in}^{(CR)} \otimes \rho_{in}^{(CTC)} \right) U^{\dagger} \right].
\end{equation}
The output state $\rho_{out}^{(CR)}$ is in general non-unitary evolution of  $\rho_{in}^{(CR)}$, because  $\rho_{out}^{(CR)}$ depends on both 
$\rho_{in}^{(CR)}$ and $\rho^{(CTC)}$, and $\rho^{(CTC)}$ also depends on  $\rho_{in}^{(CR)}$.

The post-selected CTCs\cite{p-ctc0,p-ctc1,p-ctc2} (P-CTCs) are  another type of quantum mechanical CTCs, which also solve the paradoxes. 
In P-CTC picture time travel effectively represents a quantum communication channel from the future to the past. If one chooses a quantum teleportation
channel as the communication channel, P-CTCs provide a 
self-consistent picture of quantum mechanical time travel via post-selected quantum teleportation\cite{teleportation}. It is based on the Horowitz-Maldacena
``final state condition''\cite{horo04} for black hole evaporation\cite{hawking1} and, unlike D-CTCs, are consistent with path-integral approaches to 
CTCs\cite{politzer92,politzer94}. In P-CTCs formalism the state in CTC-system is not explicitly specified while the output state of the CR-system is 
given by 
\begin{equation}
\label{CRevolution2}
\rho_{out}^{(CR)} \propto V \rho_{in}^{(CR)} V^{\dagger}
\end{equation}
where $V = \mbox{tr}_{CTC} U$. It turned out that though P-CTCs are less powerful resource than D-CTCs in the quantum information processing, 
they also have a computational and discrimination power\cite{brun2012}.

In this Letter we explore the following issues. By introducing simple qubit system and general controlled operations mixedness of the CR system is 
examined. The mixedness is measured by a magnitude of Bloch vector. It is shown that the magnitude of Bloch vector for qubit system assisted by 
D-CTCs either preserves or decreases. Thus, the pure CR-state can propagate to mixed state when CTC-qubit travels around the D-CTC. In this sense,  CTC-problem resembles the information 
loss problem\cite{loss1,loss2} in Hawking radiation\cite{ralph07}. 
For P-CTCs, however, the magnitude of Bloch vector can increase. In this case a mixed state
can evolve to a pure state. Even the completely mixed state can be converted into a pure state if the controlled operation is chosen appropriately. 
We also examine how the entanglement of the CR-system is changed in the presence of CTCs. While D-CTCs always either preserve or degrade 
the entanglement, P-CTCs can increase it. Thus, if any partially entangled CR-state is prepared, one can change it into a maximally entangled pure state 
if P-CTCs assist. This fact implies that distillation of entanglement\cite{purification1,purification2} can be easily achieved without preparing the many 
copies of the partially entangled state if P-CTCs are appropriately exploited.

\begin{figure}[ht!]
\begin{center}
\includegraphics[height=4.5cm]{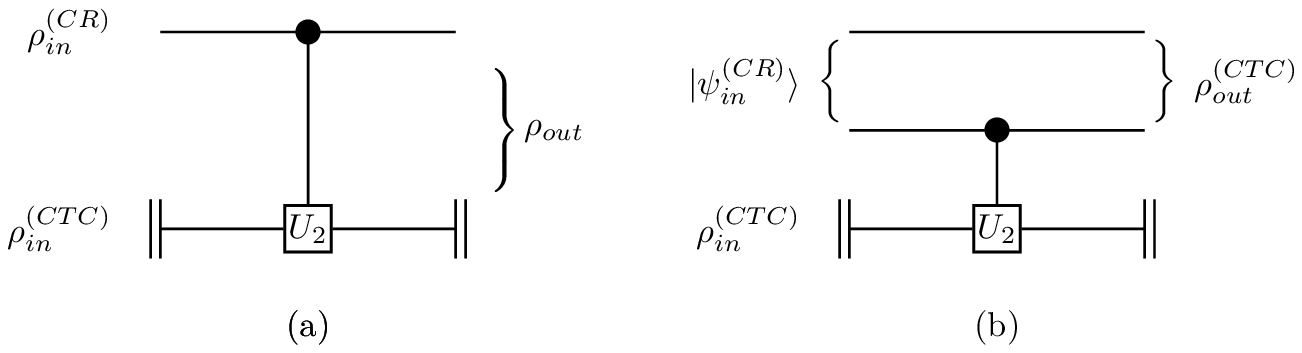}
\caption[fig1]{(a) Circuit for examining the mixedness of CR-system when CR and CTC systems interact with each other through general controlled 
operations.
The $U_2$ is represented by Eq. (\ref{defU2}). The double vertical bars on the bottom left and right indicate the past and future mouths of the wormhole 
for the CTC. (b) Circuit for examining the entanglement of CR-state in the presence of CTC. We choose the initial CR-state as a partially entangled 
state $\ket{\psi_{in}^{(CR)}} = \alpha \ket{00} + \beta \ket{11}$ with $\alpha^2 + \beta^2 = 1$ and $|\beta| \geq |\alpha|$.}
\end{center}
\end{figure}
%%%%%%%%%%%%%%%%%%%%%%%%%%%%%%%%%%%%%%%%%%%%%%%%%%%%%%%%%%%%%

\section{Mixedness in the presence of CTCs}
Since both D-CTC and P-CTC allow the non-unitary evolution for CR system, it is natural to ask how the mixedness is changed in the 
presence of the CTCs. It is known that single-qubit state is pure or completely mixed when the magnitude of Bloch vector is one or zero.
Thus, it is natural to choose the magnitude of Bloch vector as a measure of mixedness.

 The most general interaction between single-qubit CR and CTC systems is unitary group $U(2)$, whose generators are Pauli matrices and identity.
 Treating the general $U_2$ interaction seems to be difficult because it has too many free parameters. Since we want to to discuss the mixedness and 
 entanglement issues on the analytic ground, we want to choose more simple interaction. Since the controlled-gate is frequently used in the quantum
 information processing, we choose the controlled-$U_2$ interaction between CR and CTC systems. In this case 
$U_2$ is represented by four real parameters as follows;
\begin{eqnarray}
\label{defU2}
U_2 = e^{i \phi /2} \left(      \begin{array}{cc}
                                     \cos \theta e^{i \phi_1}  &  \sin \theta e^{i \phi_2}    \\
                                     -\sin \theta e^{-i \phi_2}  &  \cos \theta e^{-i \phi_1}
                                          \end{array}                                                                 \right).
\end{eqnarray}
The initial CR-state is chosen as a general form of one-qubit $\rho_{in}^{(CR)} = \frac{1}{2} \left( I_2 + {\bm r} \cdot {\bm \sigma} \right)$, 
where $|{\bm r}| = 0$ and $|{\bm r}| = 1$ correspond to the completely mixed and pure states, respectively. We assume $r_3 \neq 1$ because if 
$r_3=1$, the controlled operation cannot be turned on.

For the case of P-CTC one can derive the output-CR state by making use of Eq. (\ref{CRevolution2}) as  
$ \rho_{out}^{CR} = \frac{1}{2} (I_2 + {\bm r'} \cdot {\bm \sigma})$, where
\begin{eqnarray}
\label{pctc-2}
& &r_1' = \frac{2 \cos \theta \cos \phi_1}{(1 + r_3) + \cos^2 \theta \cos^2 \phi_1 (1 - r_3)} \left( r_1 \cos \frac{\phi}{2} - r_2  \sin \frac{\phi}{2} \right)                                                \nonumber     \\
& &r_2' = \frac{2 \cos \theta \cos \phi_1}{(1 + r_3) + \cos^2 \theta \cos^2 \phi_1 (1 - r_3)} \left( r_1 \sin \frac{\phi}{2} + r_2  \cos \frac{\phi}{2} \right)                                                \\      \nonumber
& &r_3' = \frac{(1 + r_3) - \cos^2 \theta \cos^2 \phi_1 (1 - r_3)}{(1 + r_3) + \cos^2 \theta \cos^2 \phi_1 (1 - r_3)}.   
\end{eqnarray}
Then, one can show directly
\begin{equation}
\label{pctc-3}
|{\bm r'}|^2 - |{\bm r}|^2 = (1 -  |{\bm r}|^2 ) \left[ 1 - \left( \frac{2 \cos \theta \cos \phi_1}{(1 + r_3) + \cos^2 \theta \cos^2 \phi_1 (1 - r_3)} \right)^2 \right].
\end{equation}
As expected, Eq. (\ref{pctc-3}) guarantees that the pure input CR-state always evolves into pure. Since, however, the right-hand side of Eq. (\ref{pctc-3}) can be 
positive or negative depending on $U_2$, the CR-state can evolve with increasing or decreasing its mixedness. Even though $\rho_{in}^{(CR)}$ is 
completely mixed state,  $\rho_{out}^{(CR)}$ becomes pure state $\ket{0} \bra{0}$ when $\theta = \pi / 2$ or $\phi_1 = \pi / 2$. 
Thus, P-CTC allows the evolution from mixed to pure state if qubit travels around the P-CTC. 

However, the situation is different if the CR-system is assisted by D-CTC. If the initial CTC-state  $\rho_{in}^{(CTC)}$ is chosen as 
a general form  $\rho_{in}^{(CTC)} = \frac{1}{2} \left( I_2 + {\bm s} \cdot {\bm \sigma} \right)$, one can show directly 
$\rho_{out}^{(CTC)}  \equiv \mbox{tr}_{CR} \left[U \rho_{in}^{(CR)} \otimes \rho_{in}^{(CTC)} U^{\dagger} \right]= \frac{1}{2} 
\left( I_2 + {\bm s'} \cdot {\bm \sigma} \right)$, where
\begin{eqnarray}
\label{output2}
& &\Delta s_1  = 
-(1 - r_3) \bigg[ s_1 \left(\sin^2 \phi_1 + \sin^2 \theta \cos(\phi_1 + \phi_2) \cos(\phi_1 - \phi_2)  \right)    \nonumber  \\
& &
- s_2 \left( \cos^2 \theta \sin \phi_1 \cos \phi_1 + \sin^2 \theta \sin \phi_2 \cos \phi_2 \right) + s_3 \sin \theta \cos \theta \cos (\phi_1 + \phi_2) \bigg]                                                                                                                                                    \\   \nonumber
& &\Delta s_2 = 
-(1 - r_3) \bigg[ s_1 \left( \cos^2 \theta \sin \phi_1 \cos \phi_1 - \sin^2 \theta \sin \phi_2 \cos \phi_2 \right)      \\   \nonumber
& &
+ s_2 \left(  \sin^2 \phi_1 - \sin^2 \theta \sin(\phi_1 + \phi_2) \sin (\phi_1 - \phi_2) \right) - s_3 \sin \theta \cos \theta \sin (\phi_1 + \phi_2) \bigg]                                                                                                                                                              \\   \nonumber
& &\Delta s_3= (1 - r_3) \sin \theta \bigg[ s_1 \cos \theta \cos (\phi_1 - \phi_2) + s_2 \cos \theta  \sin (\phi_1 - \phi_2)
-s_3 \sin \theta \bigg]
\end{eqnarray}
with  $\Delta s_j = s_j' - s_j \hspace{.2cm} (j=1, 2, 3)$. Then, the self-consistency condition (\ref{self}) simply reduces to $\Delta s_j = 0$. 

\begin{center}
\begin{tabular}{c|c} \hline \hline
 condition &  solution of self-consistency condition     \\  \hline 
 $\sin \theta = 0, \sin \phi_1 = 0$ \hspace{.2cm} &  no constraint    \\
 $\sin \theta = 0, \sin \phi_1 \neq 0$ \hspace{.2cm} & $s_1 = s_2 = 0$      \\
 $ \sin \theta \neq 0, \sin \phi_1 = 0$ \hspace{.2cm} &  $s_1 = s_2 \tan \phi_2, s_3 = 0$     \\
 $ \sin \theta \neq 0, \sin \phi_1 \neq 0$ \hspace{.2cm} & \hspace{.2cm} $s_1 = s_3 \tan \theta \csc \phi_1 \sin \phi_2, s_2 = s_3 \tan \theta \csc \phi_1 \cos \phi_2$  \\ \hline   \hline
\end{tabular}

\vspace{0.1cm}
Table I: Solutions of the self-consistency condition for various $U_2$.
\end{center}

The solutions of the self-consistency condition are summarized in Table I for various $U_2$. We will focus on  the case of $\sin \theta \neq 0$
and $\sin \phi_1 \neq 0$, since the remaining ones can be discussed in a similar way. Since $|{\bm s}| \leq 1$, the self-consistency condition 
implies
\begin{equation}
\label{out2-5}
s_3^2 \leq \frac{\sin^2 \phi_1}{\sin^2 \phi_1 + \tan^2 \theta}
\end{equation}
where equality holds for pure CTC-state. Then, the output CR-state becomes $\rho_{out}^{(CR)}  \equiv \mbox{tr}_{CTC} \left[U \rho_{in}^{(CR)}
\otimes \rho_{in}^{(CTC)} U^{\dagger} \right]= \frac{1}{2} \left( I_2 + {\bm r'} \cdot {\bm \sigma} \right)$, where 
\begin{equation}
\label{out2-6}
r_1' = P r_1 - Q r_2     \hspace{1.0cm}  r_2' = Q r_1 + P r_2 \hspace{1.0cm} r_3' = r_3
\end{equation}
with
\begin{eqnarray}
\label{out2-7}
& &P = \cos \frac{\phi}{2} \cos \theta \cos \phi_1 - s_3 \sin \frac{\phi}{2} \frac{\sin^2 \theta + \cos^2 \theta \sin^2 \phi_1}{\cos \theta \sin \phi_1}              \\   \nonumber
& &Q =  \sin \frac{\phi}{2} \cos \theta \cos \phi_1 + s_3  \cos \frac{\phi}{2} \frac{\sin^2 \theta + \cos^2 \theta \sin^2 \phi_1}{\cos \theta \sin \phi_1}.
\end{eqnarray}
Therefore, $|{\bm r'}|^2 = (P^2 + Q^2) (r_1^2 + r_2^2) + r_3^2$, where 
$$P^2 + Q^2 = \cos^2 \theta \cos^2 \phi_1 + s_3^2 \left(\frac{\sin^2 \theta + \cos^2 \theta \sin^2 \phi_1)}{\cos \theta \sin \phi_1}\right)^2.$$
When $s_3^2$ saturates the inequality (\ref{out2-5}), it is easy to show $|{\bm r'}| = |{\bm r}|$. Thus, the mixedness of the CR-system is 
preserved when the CTC-system is pure. When CTC-state is mixed, $\rho_{out}^{(CR)}$ is more mixed than $\rho_{in}^{(CR)}$, i.e.
$|{\bm r'}| < |{\bm r}|$. If the Deutsch's maximum entropy postulate is chosen,  $\rho_{out}^{(CR)}$ becomes the maximal mixed state
$|{\bm r'}|^2 = \cos^2 \theta \cos^2 \phi_1 (r_1^2 + r_2^2) + r_3^2$. Thus, any pure states of the form 
$\frac{1}{\sqrt{2}} \left( \ket{0} + e^{i \theta} \ket{1} \right)$ can be converted into the completely mixed state when $\cos \theta =0$ or 
$\cos \phi_1 = 0$ if maximum entropy rule is chosen.

\section{Entanglement in the presence of CTCs}
We examine how the entanglement of CR-system is changed in the presence of CTCs. To explore this issue we introduce partially 
entangled two-qubit initial state $\ket{\psi_{in}^{(CR)}} = \alpha \ket{00} + \beta \ket{11}$ where $\alpha^2 + \beta^2 = 1$. We also choose
$|\beta| \geq |\alpha|$ without loss of generality. One party of CR-system interacts with CTC through the controlled-$U_2$ operation. The other party
has no interaction with the CTC-system. This situation is depicted in Fig. 1(b) as a quantum circuit. We will use the concurrence\cite{woot-98} as an
entanglement measure. The concurrence of $\ket{\psi_{in}^{(CR)}}$ is $2 |\alpha \beta|$.

%%%%%%%%%%%%%%%%%%%%%%%%%%%%%%%%%%%%%%%%%%%%%%%%%%%%%%
%\begin{figure}[ht!]
%\begin{center}
%\includegraphics[height=4.0cm]{fig2.eps}
%\includegraphics[height=6.0cm]{fig7b.eps}
%\includegraphics[height=6.0cm]{fig7c.eps}
%\caption[fig2]{XXXXXXXXXXXXXXXXXXXXXXXXXXXXX}
%\end{center}
%\end{figure}
%%%%%%%%%%%%%%%%%%%%%%%%%%%%%%%%%%%%%%%%%%%%%%%%%%%%%%%%%%%%%

For the case of P-CTC one can derive $\rho_{out}^{(CR)}$ by making use of Eq. (\ref{CRevolution2}) in a form
\begin{eqnarray}
\label{pctc-5}
& &\rho_{out}^{(CR)} = \frac{1}{\alpha^2 + \beta^2 \cos^2 \theta \cos^2 \phi_1}
\bigg[ \alpha^2 \ket{00} \bra{00} + \beta^2 \cos^2 \theta \cos^2 \phi_1 \ket{11} \bra{11}     \\   \nonumber
& &\hspace{3.0cm}   + \alpha \beta e^{-i \phi / 2} \cos \theta \cos \phi_1 \ket{00} \bra{11} 
+ \alpha \beta e^{i \phi / 2} \cos \theta \cos \phi_1 \ket{11} \bra{00} \bigg].
\end{eqnarray} 
The concurrence of $\rho_{out}^{(CR)}$ is easily computed by following the procedure of Ref.\cite{woot-98} and final expression is 
\begin{equation}
\label{pctc-6}
{\cal C} \left( \rho_{out}^{(CR)} \right) = 2 |\alpha \beta| \gamma
\end{equation}
where the ratio $\gamma$ is 
\begin{equation}
\label{pctc-7}
\gamma = \frac{|\cos \theta \cos \phi_1|}{\alpha^2 + \beta^2 \cos^2 \theta \cos^2 \phi_1}.
\end{equation}
It is remarkable to note that the ratio $\gamma$ is dependent on both $U_2$ and the initial CR-state.  
%${\cal C} \left( \rho_{out}^{(CR)} \right)$ is plotted in Fig. 2 as a function of $\alpha$ and $\cos \theta \cos \phi_1$. 
Surprisingly, one can always make 
$ \rho_{out}^{(CR)}$ maximally entangled pure state $\frac{1}{\sqrt{2}} (\ket{00} \pm e^{i \phi / 2} \ket{11})$ by choosing $\cos \theta \cos \phi_1 = \pm \frac{\alpha}{\beta}$. Thus, if P-CTC exists, the 
distillation of entanglement of CR-system can be easily achieved without preparing multiple copies of the partially entangled state. It is sufficient to 
prepare a single copy for complete distillation by choosing $U_2$ appropriately.

The situation is different for the case of D-CTC. We define the initial CTC-state as a one-qubit general form 
$\rho_{in}^{(CTC)} = \frac{1}{2} \left( I_2 + {\bm s} \cdot {\bm \sigma} \right)$. Then, the output CTC state becomes
$\rho_{out}^{(CTC)} \equiv \mbox{tr}_{CR} \left[ U \ket{\psi_{in}^{(CR)}}\bra{\psi_{in}^{(CR)}}\otimes \rho_{in}^{(CTC)}U^{\dagger} \right] =  \frac{1}{2} \left( I_2 + {\bm s'} \cdot {\bm \sigma} \right)$,
where $\Delta s_j \hspace{.2cm} (j=1, 2, 3)$ are exactly the same with Eq. (\ref{output2}) if $1 - r_3$ is changed into $2 \beta^2$. Thus, the solutions of
the self-consistency condition are identical with those given in Table I. One can also show directly that the output CR-state is
\begin{equation}
\label{out3-1}
\rho_{out}^{(CR)} \equiv \mbox{tr}_{CTC} \left[ U \ket{\psi_{in}^{(CR)}}\bra{\psi_{in}^{(CR)}}\otimes \rho_{in}^{(CTC)}U^{\dagger} \right]
=\alpha^2 \ket{00}\bra{00} + \beta^2 \ket{11} \bra{11} + A \ket{00} \bra{11} + A^* \ket{11} \bra{00}
\end{equation}
where
\begin{equation}
\label{out3-2}
A = e^{-i \phi /2} \alpha \beta \left[ \cos \theta \cos \phi_1 - i \left( s_1 \sin \theta  \sin \phi_2 + s_2 \sin \theta \cos \phi_2 + s_3 
\cos \theta \sin \phi_1 \right) \right].
\end{equation}
It is easy to show that the concurrence of $\rho_{out}^{(CR)}$ is 
\begin{equation}
\label{out3-3}
{\cal C} \left(\rho_{out}^{(CR)} \right) = 2 \min \left( |A|, |\alpha \beta| \right).
\end{equation}
Thus, D-CTC can either preserve or decrease the entanglement of CR-system.

For example, let us consider the case of $\sin \theta \neq 0$ and $\sin \phi_1 \neq 0$. Then, the variation of entanglement 
$\Delta {\cal E} \equiv {\cal C} \left(\rho_{in}^{(CR)} \right) - {\cal C} \left(\rho_{out}^{(CR)} \right)$ can be computed by making use of 
Eq. (\ref{out3-3}) and Table I:
\begin{equation}
\label{out3-8}
\Delta {\cal E} = 2 |\alpha \beta| \left[ 1 - \sqrt{1 - \left( 1 - \frac{\sin^2 \phi_1 + \tan^2 \theta}{\sin^2 \phi_1} s_3^2 \right) 
\left( \sin^2 \theta + \cos^2 \theta \sin^2 \phi_1 \right)} \right].
\end{equation}
Thus, if the inequality (\ref{out2-5}) is saturated, $\Delta {\cal E}$ vanishes. This means that if the CTC-state is pure, the entanglement of CR-state is 
preserved. If we choose the maximal entropy CTC-state as Deutsch suggested, the maximal degradation of entanglement
$\Delta  {\cal E} = 2 |\alpha \beta| (1 - |\cos \theta \cos \phi_1|)$ occurs.

\section{Conclusions}
Although the theory of general relativity does allow CTC as a solution of Einstein field equations, still there are a lot of controversial for existence of 
CTCs. In this Letter we have addressed two issues, mixedness and entanglement for CR system assuming that D-CTC and/or P-CTC exist(s) in 
nature. It was shown that while D-CTC-assisted qubit cannot increase the magnitude of its Bloch vector, P-CTC-assisted qubit can. As a result, the mixed CR-state can evolve to pure CR-state if P-CTC exists. Even the completely mixed state can evolve to pure state if we choose the phase angles of $U_2$ 
appropriately.

Although the CTC-state is not specified explicitly for the case of P-CTC, one can get some information of P-CTC-state if exists any. Let us imagine a 
closed systems composed by CR and P-CTC subsystems. Let us assume that they interacts with each other through some unitary operation. If one uses
the subadditivity of the von Neumann entropy one can show $\Delta S^{(CTC)} \geq -\Delta S^{(CR)}$, where $S$ is a von Neumann entropy and 
$\Delta S^{(\cdot)} \equiv S \left(\rho_{out}^{(\cdot)}\right) - S \left(\rho_{in}^{(\cdot)}\right)$. Thus, computing the entropy difference of 
CR-subsystem one can compute the lower bound of  $\Delta S^{(CTC)}$ although we do not know the P-CTC state explicitly.

We also have studied the case where the CR-system consists of bipartite partially entangled particles and one of them interacts with CTC system 
through controlled-$U_2$ operation. For the case of P-CTC surprisingly the partially entangled state can always be converted into the  
maximally entangled 
pure state if the phase angles of $U_2$ are chosen appropriately. If, therefore, P-CTCs exist, the distillation protocol of entanglement
is easily achieved without preparing the multiple copies of the partially entangled state. For the case of D-CTC such a nonintuitive effect disappears because 
D-CTC either preserves or decreases the entanglement of CR system.

There are a lot of questions in the context of CTCs. How to incorporate the general relativistic CTCs into the quantum mechanical CTCs or {\it vice versa}? What happens to the uncertainty relations if CTCs exist\cite{brun13,pienaar13,yuan14}? While P-CTC allows only the evolution of pure state to pure state, 
D-CTC allow the evolution of pure state to mixed state. Thus existence of D-CTC may provide clue for the information loss problem of black hole.
In this context the information loss problem was discussed in Ref.\cite{ralph07} by introducing a simple toy model.
Of course, rigorous 
and explicit analysis should be addressed to clarify a connection between existence of CTCs and information loss problem.
Probably, the theory of quantum gravity may give some answers in the future.

Another interesting issue is to check whether our results of mixedness and entanglement in the presence of CTCs are valid for general 
interaction $U(2)$ or not. We hope to study this issue in the future.

\begin{acknowledgments}
This research was supported by the Basic Science Research Program through the National Research Foundation of Korea(NRF) funded by the Ministry of Education, Science and Technology(2011-0011971).
\end{acknowledgments}

\end{document}